# Binary classification models with "Uncertain" predictions

Damjan Krstajic[1§], Ljubomir Buturovic[2], Simon Thomas[3], David E Leahy[4]

[1]Research Centre for Cheminformatics, Jasenova 7, 11030 Beograd, Serbia

[2]Clinical Persona, 932 Mouton Circle, East Palo Alto, CA 94303, USA

[3]Cyprotex Discovery Ltd, No. 24 Mereside, Alderley Park, Macclesfield

SK10 4TG, UK

[4]Discovery Bus Ltd, The Oakridge Centre, Gibhill Farm, Shrigley Rd, Macclesfield,

SK10 5SE, UK

[§]Corresponding author

Email addresses:

    DK: damjan.krstajic@rcc.org.rs

    LB: ljubomir@clinicalpersona.com

    ST: s.thomas@cyprotex.com

    DEL: david.leahy@discoverybus.com

# Abstract

Binary classification models which can assign probabilities to categories such as "the tissue is 75% likely to be tumorous" or "the chemical is 25% likely to be toxic" are well understood statistically, but their utility as an input to decision making is less well explored. We argue that users need to know which is the most probable outcome, how likely that is to be true and, in addition, whether the model is capable enough to provide an answer. It is the last case, where the potential outcomes of the model explicitly include 'don't know' that is addressed in this paper. Including this outcome would better separate those predictions that can lead directly to a decision from those where more data is needed. Where models produce an "Uncertain" answer similar to a human reply of "don't know" or "50:50" in the examples we refer to earlier, this would translate to actions such as "operate on tumour" or "remove compound from use" where the models give a "more true than not" answer. Where the models judge the result "Uncertain" the practical decision might be "carry out more detailed laboratory testing of compound" or "commission new tissue analyses". The paper presents several examples where we first analyse the effect of its introduction, then present a methodology for separating "Uncertain" from binary predictions and finally, we provide arguments for its use in practice.

## Background

In our practice we have come up with the following situation more than once. We create a binary classification model on a training dataset and predict a test dataset. We again execute the same process of building a classification model and predicting the test dataset on the same machine, and we find that one or more test samples have the opposite predicted categories. How is this possible? We use the same training dataset and create the binary model with exactly the same input parameters on the same machine. Upon further investigation we find that the problematic samples have predicted probabilities close to 0.5. Due to random choice inside the algorithm of our model building technique, even though the inputs were the same and the machine was the same, the output equations of the models were not exactly the same. Therefore, the predicted probabilities of the problematic test samples were sometimes above 0.5 and sometimes below, thus producing opposite predicted categories. So how should we report the predicted categories of the problematic test samples?

An obvious answer would be to report them as "Uncertain". However, even if we were to report them as "Uncertain", how would we in future, using the same binary classification model, differentiate between "Uncertain" and the two categories?

# Introduction

First, we would like to clarify that a binary classification prediction of probability for a test sample equal to 0.5 could mean different things. It could mean that the test sample is so different from our training dataset that there are not any arguments for the model to classify it as belonging to any of the two categories. Or, it could be that the model recognises equally large numbers of arguments in favour and against classifying it to any of the two categories. In the former instance a human reply would be equal to "don't know", while for the latter it would be "50:50" [1]. Regardless of this subtle and important difference, we think that we should allow binary classification models to have a third prediction category called "Uncertain", which could mean either "don't know" or "50:50".

There is also an argument based on psychological research for introducing a third prediction category for binary classification models which is related to the way individuals interpret or make use of probability estimates. Tversky and Kahneman [2] [3] found that most people could not keep a consistent view of what different numerical probabilities meant, and therefore "anchor", i.e. rely too much on the first piece of information offered to them when making decisions. The best they could find was that people could keep a consistent sense of the meaning of 50:50 and the meaning of "almost certain". This means that a decision maker who tries to distinguish between a 0.85 probability and a 0.7 probability of a predicted category cannot really tell the difference and in practice the human decision maker is unlikely to give any significance to the difference.

So if we predict that a compound is toxic with probability 0.89, then does it make any difference to a toxicologist if it is 0.83? Similarly, if we predict that a patient has an aggressive cancer with probability 0.29, then does it make any difference to an oncologist if it is 0.31? According to Tversky and Kahneman [2][3] anchoring may effect both the toxicologist and oncologist. However, it appears that it would not be the same if the predicted probability is close to 0.5. We would like to emphasise here that we are here only concerned with individual probability predictions. We are not underestimating in any way the importance of probability predictions when comparing different potential toxic compounds or different patients, especially when they are well calibrated [4].

Tversky and Kahneman's [2][3] findings lead Suppes to propose a simple probabilistic model with only five probabilities (described in [5]):
1. surely true
2. more probable than not
3. as probable as not
4. less probable than not
5. surely false

As we cannot have "surely true" and "surely false" when predicting unknown test samples, we propose that in binary classification settings predictive models generate the following three predictive categories to decision makers:
- Positive – "more probable than not"
- Uncertain - "as probable as not"
- Negative - "less probable than not"

We have shown that in practice we may have a situation where one or more test samples have predicted probabilities above or below 0.5 depending on the random number used by the model building algorithm. We argue that those test samples ought to be categorised as Uncertain. However, we first need to analyse the effects of having Uncertain predictions. For example, if we were to introduce it as an additional predicted category for binary classification models, would the performance of non-Uncertain predictions be improved? If not, then there is no point in introducing it. Later we will suggest an approach to defining Uncertain predictions. We will also examine the introduction of Uncertain predictions when a decision threshold is different from 0.5 in order to optimise specificity and sensitivity. Finally we will discuss the implications and benefits of having Uncertain predictions in practice.

## Methods

**Analysing and selecting the interval of uncertainty**

Our analysis of Uncertain predictions for binary classification models relies on the ability of the binary model to produce the probability of a predicted category and is not applicable to binary classification models which are not able to produce probabilities of their predicted categories.

The first step is to check whether the introduction of Uncertain predictions would reduce the misclassification error of a classifier. Therefore, we are interested in assessing the misclassification error for cases when all predicted probabilities in the range between 0.49 and 0.51 are defined as Uncertain. Similarly, we are interested to perform the same analysis for other intervals such as (0.48, 0.52), (0.47,0.53), etc.

We use repeated grid-search cross-validation [6] for model selection as well as for analysing and defining Uncertain categories. For each repeated cross-validation we first define Uncertain predictions for all cross-validated probabilities in the interval (0.49, 0.51) and calculate misclassification error for non-Uncertain predictions as well as the proportion of Uncertain predictions. Using results from all repeated cross-validations we then calculate the mean misclassification error as well as the mean proportion of Uncertain predictions. We repeat the same process for other intervals (0.48, 0.52), (0.47, 0.53), etc. We refer to the interval of probabilities which is associated with the Uncertain prediction as the *interval of uncertainty*.

If the analysis confirms that the introduction of Uncertain predictions improves the performance of the classifier for non-Uncertain predictions then one may select an interval of uncertainty which would then define the Uncertain predictive category. One may say that for example 20% of Uncertain predictions is acceptable and select an interval of uncertainty which has the mean of proportion of Uncertain predictions close to 20%. However, we are more interested in a systematic approach to defining Uncertain predictions.

**Collection of N binary classification models**

In the Background section we gave an example of test compounds whose predictions flip-flop between two opposite categories depending on the random number used by the model building method. It is not feasible to create models with various random numbers in order to find predictions which flip-flop between opposite categories. One way to simulate something similar is to create models on training datasets which differ in one sample only. If a training dataset consists of N samples then we could create N different binary classification models on (N-1) samples and for each test sample provide N predicted categories. We would then define an Uncertain prediction to be a test sample which has opposite predicted categories among N predictions. This would mean that instead of applying a single binary classification model in practice we would need to execute a collection of N binary classification models.

In order to assess the impact of such approach we will execute repeated grid-search cross-validation [6], where in addition to building a probabilistic binary classification model on a learning dataset we would also create probabilistic binary classification models on all possible subsets of the learning dataset with one sample missing. If any two models built on subsets of the learning dataset predicted opposite categories,we would categorise it as Uncertain. At the end of the repeated cross-validation process we would be able to analyse which probability predictions from the model built on the learning datasets were categorised as Uncertain using the collection of N binary classification models, as well as check if the introduction of Uncertain predictions reduced the misclassification error. Using predicted probabilities from N binary classification models we will also calculate minimum, mean and maximum probability for each validation sample.

**Classification models with a decision threshold different from 0.5**

Sometimes a decision threshold different from 0.5 is chosen in order to optimise classification measures such as specificity and sensitivity. We are interested to assess the specificity and sensitivity for cases when all predicted probabilities in the range +/- 0.01 from the decision threshold are defined as Uncertain. Similarly, we are interested to perform the same analysis for other intervals, such as +/- 0.02 from the decision threshold, +/- 0.03 from the decision threshold, etc.

## Materials

We used three datasets from the QSARdata R package [7] with binary outcomes to demonstrate our methods:

- *bbb2* contains 80 compounds, 45 categorised as "crossing", the other 35 categorised as "not crossing". We used LCALC descriptors because when compared to other descriptor sets it generated better models (results not shown). We removed chloramphenicol from the dataset because LCALC descriptors were not provided for it. During pre-processing [8][9] we removed descriptor LCALC_NDA as it was a linear combinations of the remaining 22 descriptors.
- *Mutagen* contains 4335 compounds, 2400 categorised as "mutagen", the other 1935 compounds as "nonmutagen". During pre-processing [8][9] we removed 281 descriptors with near zero variation and 15 descriptors that were linear combinations of others, thus leaving 1283 descriptors for model building.

- *PLD* contains 324 compounds, 124 categorised as "inducer", the other 200 as "noninducer". We used PipelinePilotFP descriptors because when compared to other descriptor sets it generated better models (results not shown). During pre-processing [8][9] we removed 2183 descriptors with near zero variation and 371 descriptors that were linear combinations of others, thus leaving 308 descriptors for model building.

As regards classification model building techniques we applied the following two:
- ridge logistic regression – We applied ridge logistic regression [10] using the glmnet R package [11][12]. We let the glmnet function compute its own array of lambda values based on nlambda = 100 and lambda.min.ratio = $10^{-6}$.
- random forest – We applied random forest [13] using the randomForest R package [14][15]. We used all default input parameters, i.e. without any grid-search.

# Results

**Analysing the interval of uncertainty**

We analysed the introduction of Uncertain predictions on three datasets with two binary classification models. We executed 50 times 2-fold cross-validation and selected models with the least mean misclassification error for each dataset. We chose 2-fold cross-validation because we were interested to have as many test samples as possible. In Tables 1, 2, 3 we show mean misclassification error and mean percentage of Uncertain predictions for each interval of uncertainty. In all examples we can confirm that the greater the interval of probabilities for Uncertain predictions is, the lower the mean misclassification error. However, from the results in Tables 1, 2, 3 it is not obvious where an optimal cut-off for Uncertain predictions would be.

**Collection of N binary classification models**

Based on the results from our above analysis we selected the models with the least mean classification error, and we applied 50 times repeated 2-fold cross-validation, where we created a probabilistic classification model on a learning dataset, as well as classification models on all available subsets of the learning dataset with one sample missing. We categorised test samples with opposite predictions from any two models built on subsets of the learning dataset as Uncertain. In Table 4 we show the mean misclassification error and mean percentage of Uncertain predictions for each pair of model and dataset.

We were interested to see if we could define an interval of uncertainty using repeated cross-validation with a collection of N binary classification models. Unfortunately, we have found that certain validation samples may have similar predicted probabilities based on a model built on a learning dataset, while predicted probabilities from N binary classification models may vary differently. In Table 5 we present some examples of pairs of predictions which show contradiction. For example we built a random forest model on half of the PLD dataset and the predicted probabilities for Memantine and Indoramin are 0.7 and 0.6 respectively. However, the minimum and maximum ranges for predicted probabilities from the collection of N binary classification models for Memantine and Indoramin are (0.42, 0.752) and (0.54, 0.67) respectively. This means that even though the predicted probability of Indoramin (p=0.6) is much closer to 0.5 than Memantine (p=0.7), the collection of N models classified Memantine as an Uncertain prediction, while Indoramin was predicted not to be. The contradictory cases like the ones shown in Table 5 are not rare in our examples. Furthermore, it appears that the collection of N binary classification models generates some kind of "personalised" range of probabilities for each validation sample.

**Classification models with a decision threshold different from 0.5**

We were interested in examining whether the introduction of Uncertain predictions may improve both specificity and sensitivity when a decision threshold is different from 0.5. We chose two different decision thresholds (0.3 and 0.7) in our analysis. In Tables 6a, 6b, 7a, 7b, 8a, 8b we show mean specificity, mean sensitivity and mean proportion of Uncertain predictions for each interval of uncertainty. In no example did we have a case where introduction of Uncertain predictions improved both specificity and sensitivity.

## Discussion

Our starting research question was: If human experts may say "I don't know" to a valid question, why then do statistical models and expert software systems not have that answer built in them? It has been found that in intuitionistic fuzzy sets [16][17] it is possible to have a situation where the sum of probabilities of opposite events is less than 1. So we may say that the probability of a person having an aggressive cancer is 0.3 and the probability of not having the aggressive cancer is 0.25. In that case the degree of uncertainty is 0.45 = 1 – 0.3 – 0.25. There is active research into combining statistical models with intuitionistic fuzzy logic [18][19]. However, as far as we are aware, currently all binary classification models in practice provide a probability P of belonging to a category, which means (1-P) probability of belonging to the opposite category. Therefore we are currently unable to differentiate between 50:50 and "don't know" answers. Schmidt and Kreinovich [1] use the terms TRUE, FALSE and UNCERTAIN, and we have borrowed the term Uncertain from them.

When reading our Background section one may rightly argue that by fixing a random seed number we would always get the same predictions. However true this may be, it does not change the fact that certain test samples will have opposite predictions depending on the random number. We also acknowledge that the introduction of the Uncertain category has not removed a possible influence of random numbers on our predictions. We still may have test samples which flip-flop between Negative and Uncertain and between Uncertain and Positive, but that are very unlikely to flip-flop between Negative and Positive.

In all our examples the introduction of Uncertain has reduced the misclassification error. If we use a decision threshold different from 0.5 in order to optimise sensitivity and specificity, we could not see in our examples any improvements to both specificity and sensitivity when introducing Uncertain predictions. Our examples might not be representative for general use so we advise analysing the introduction of Uncertain predictions on every dataset regardless of whether the decision threshold is 0.5 or not.

We think that benefits of introducing Uncertain as a predicted category in practice may outweigh any additional confusion it may generate. It is related to how we use and trust binary classification models. For example, in the case of a predictive model for an aggressive cancer, a patient with an Uncertain prognosis may provide the basis for an oncologist to ask for additional tests; while for toxicologists all compounds with Uncertain predictions would automatically be sent to the laboratory for *in vitro* screening. One may even imagine a drug discovery process in which the starting point consists of predictive models and the majority of compounds selected for *in vitro* assay would be those with Uncertain predictions.

There are other possible ways in which to define Uncertain predictions, and we are not suggesting that our approaches are the best or the only one. However, our main intention has been to provide a case for an introduction of Uncertain predictions and to propose a solution for their use.

## Conclusions

We will end with a quote from Albert Einstein: "A*s our circle of knowledge expands, so does the circumference of darkness surrounding it*". We think that the use of predictive modelling in decision making is very dependent on defining "*the circumference of darkness*", i.e. Uncertain predictions. The paper develops a rigorous approach to assessing the level of uncertainty and those cases where a model cannot identify a clear decision and where further investigation should be targeted.

# Tables

**Table 1 - Analysis of intervals of uncertainty for the Mutagen dataset**

| interval of uncertainty | ridge logistic regression | | random forest | |
|---|---|---|---|---|
| | misclassification error | percent Uncertain | misclassification error | percent Uncertain |
| - | 0.201 | 0.00 | 0.192 | 0.00 |
| (0.49 – 0.51) | 0.197 | 1.14 | 0.186 | 2.14 |
| (0.48 – 0.52) | 0.194 | 2.29 | 0.179 | 4.50 |
| (0.47 – 0.53) | 0.191 | 3.44 | 0.172 | 6.84 |
| (0.46 – 0.54) | 0.187 | 4.67 | 0.166 | 9.18 |
| (0.45 – 0.55) | 0.184 | 5.91 | 0.159 | 11.53 |
| (0.44 – 0.56) | 0.181 | 7.09 | 0.154 | 13.77 |
| (0.43 – 0.57) | 0.177 | 8.30 | 0.148 | 16.30 |
| (0.42 – 0.58) | 0.174 | 9.57 | 0.142 | 18.61 |
| (0.41 – 0.59) | 0.170 | 10.82 | 0.137 | 20.67 |
| (0.40 – 0.60) | 0.167 | 12.07 | 0.132 | 22.99 |
| (0.39 – 0.61) | 0.163 | 13.38 | 0.127 | 25.36 |
| (0.38 – 0.62) | 0.160 | 14.67 | 0.122 | 27.62 |
| (0.37 – 0.63) | 0.157 | 16.00 | 0.117 | 29.92 |
| (0.36 – 0.64) | 0.153 | 17.30 | 0.113 | 32.22 |
| (0.35 – 0.65) | 0.150 | 18.64 | 0.109 | 34.54 |
| (0.34 – 0.66) | 0.147 | 19.99 | 0.105 | 37.09 |
| (0.33 – 0.67) | 0.144 | 21.33 | 0.102 | 39.32 |
| (0.32 – 0.68) | 0.141 | 22.70 | 0.098 | 41.58 |
| (0.31 – 0.69) | 0.138 | 24.04 | 0.095 | 43.76 |
| (0.30 – 0.70) | 0.135 | 25.51 | 0.092 | 45.93 |

Mean misclassification error without Uncertain predictions and mean percentage of Uncertain predictions are shown for the best logistic regression and random forest model on the Mutagen dataset.

**Table 2 - Analysis of intervals of uncertainty for the PLD dataset**

| interval of uncertainty | ridge logistic regression | | random forest | |
|---|---|---|---|---|
| | misclassification error | percent Uncertain | misclassification error | percent Uncertain |
| - | 0.191 | 0.00 | 0.201 | 0.00 |
| (0.49 – 0.51) | 0.184 | 2.27 | 0.195 | 1.89 |
| (0.48 – 0.52) | 0.179 | 4.35 | 0.189 | 4.06 |
| (0.47 – 0.53) | 0.173 | 6.60 | 0.183 | 6.16 |
| (0.46 – 0.54) | 0.167 | 8.87 | 0.178 | 8.14 |
| (0.45 – 0.55) | 0.161 | 11.24 | 0.170 | 10.38 |
| (0.44 – 0.56) | 0.156 | 13.61 | 0.165 | 12.36 |
| (0.43 – 0.57) | 0.148 | 16.00 | 0.158 | 14.68 |
| (0.42 – 0.58) | 0.141 | 18.35 | 0.152 | 16.73 |
| (0.41 – 0.59) | 0.137 | 20.83 | 0.148 | 18.40 |
| (0.40 – 0.60) | 0.133 | 23.33 | 0.144 | 20.33 |
| (0.39 – 0.61) | 0.127 | 25.75 | 0.138 | 22.44 |
| (0.38 – 0.62) | 0.121 | 28.20 | 0.133 | 24.33 |
| (0.37 – 0.63) | 0.116 | 30.54 | 0.129 | 26.28 |
| (0.36 – 0.64) | 0.111 | 33.00 | 0.125 | 28.27 |
| (0.35 – 0.65) | 0.107 | 35.41 | 0.121 | 30.15 |
| (0.34 – 0.66) | 0.102 | 37.52 | 0.117 | 32.65 |
| (0.33 – 0.67) | 0.098 | 39.63 | 0.113 | 34.80 |
| (0.32 – 0.68) | 0.095 | 41.75 | 0.109 | 36.90 |
| (0.31 – 0.69) | 0.094 | 43.94 | 0.107 | 38.62 |
| (0.30 – 0.70) | 0.091 | 46.30 | 0.103 | 40.57 |

Mean misclassification error without Uncertain predictions and mean percentage of Uncertain predictions are shown for the best logistic regression and random forest model on the PLD dataset.

**Table 3 - Analysis of intervals of uncertainty for the bbb2 dataset**

| interval of uncertainty | ridge logistic regression | | random forest | |
|---|---|---|---|---|
| | misclassification error | percent Uncertain | misclassification error | percent Uncertain |
| - | 0.189 | 0.00 | 0.180 | 0.00 |
| (0.49 – 0.51) | 0.185 | 2.38 | 0.176 | 1.42 |
| (0.48 – 0.52) | 0.179 | 4.63 | 0.173 | 2.99 |
| (0.47 – 0.53) | 0.174 | 6.81 | 0.169 | 4.43 |
| (0.46 – 0.54) | 0.169 | 9.11 | 0.165 | 5.92 |
| (0.45 – 0.55) | 0.161 | 11.44 | 0.163 | 7.59 |
| (0.44 – 0.56) | 0.155 | 13.95 | 0.156 | 9.37 |
| (0.43 – 0.57) | 0.151 | 16.28 | 0.150 | 11.16 |
| (0.42 – 0.58) | 0.145 | 18.61 | 0.146 | 12.94 |
| (0.41 – 0.59) | 0.143 | 20.71 | 0.142 | 14.41 |
| (0.40 – 0.60) | 0.141 | 22.86 | 0.137 | 16.43 |
| (0.39 – 0.61) | 0.140 | 24.76 | 0.134 | 18.46 |
| (0.38 – 0.62) | 0.136 | 27.14 | 0.131 | 20.30 |
| (0.37 – 0.63) | 0.133 | 28.73 | 0.128 | 22.10 |
| (0.36 – 0.64) | 0.131 | 30.96 | 0.125 | 23.75 |
| (0.35 – 0.65) | 0.130 | 33.59 | 0.122 | 25.90 |
| (0.34 – 0.66) | 0.130 | 35.70 | 0.118 | 27.80 |
| (0.33 – 0.67) | 0.130 | 37.67 | 0.114 | 30.00 |
| (0.32 – 0.68) | 0.129 | 40.28 | 0.113 | 31.59 |
| (0.31 – 0.69) | 0.128 | 42.25 | 0.110 | 33.57 |
| (0.30 – 0.70) | 0.126 | 44.86 | 0.109 | 35.54 |

Mean misclassification error without Uncertain predictions and mean percentage of Uncertain predictions are shown for the best logistic regression and random forest model on the bbb2 dataset.

**Table 4 - Analysis of Uncertain predictions with collection of N binary classification models**

| dataset | ridge logistic regression | | random forest | |
|---|---|---|---|---|
| | misclassification error | percent Uncertain | misclassification error | percent Uncertain |
| bbb2 | 0.117 | 18.78 | 0.117 | 17.72 |
| PLD | 0.155 | 8.77 | 0.126 | 18.25 |
| Mutagen | 0.174 | 6.59 | 0.112 | 19.73 |

Mean misclassification error without Uncertain predictions and mean percentage of Uncertain predictions are shown for the best ridge logistic regression and random forest models on all three datasets when a collection of N binary classification models was used to define Uncertain predictions.

**Table 5 - pairs of compounds which show conflicts between closeness to 0.5 of a predicted probability generated by a model on a learning dataset and ranges of probabilities generated by a collection of N binary classification models**

| dataset | model | sample name | predicted probability | N binary classification models | | |
|---|---|---|---|---|---|---|
| | | | | min Prob | mean Prob | max Prob |
| bbb2 | random forest | Procaine | 0.604 | 0.566 | 0.623 | 0.704 |
| bbb2 | random forest | Heptacaine | 0.690 | 0.442 | 0.699 | 0.790 |
| bbb2 | ridge logistic regression | Cetirizine | 0.356 | 0.294 | 0.357 | 0.546 |
| bbb2 | ridge logistic regression | Idoxuridine | 0.414 | 0.278 | 0.411 | 0.461 |
| PLD | random forest | Memantine | 0.700 | 0.420 | 0.697 | 0.752 |
| PLD | random forest | Indoramin | 0.600 | 0.540 | 0.600 | 0.670 |
| PLD | ridge logistic regression | Fenofibrate | 0.436 | 0.408 | 0.436 | 0.503 |
| PLD | ridge logistic regression | Paraquat | 0.469 | 0.438 | 0.468 | 0.488 |
| Mutagen | random forest | 64 | 0.572 | 0.510 | 0.596 | 0.664 |
| Mutagen | random forest | 1 | 0.690 | 0.494 | 0.631 | 0.706 |
| Mutagen | ridge logistic regression | 214670 | 0.454 | 0.398 | 0.454 | 0.498 |
| Mutagen | ridge logistic regression | 214959 | 0.383 | 0.249 | 0.383 | 0.559 |

Pairs of compounds which show contradiction between the predicted probability generated by a binary classification model built on a learning dataset and predicted Unknown category with a collection of N binary classification models.

**Table 6a - Analysis of intervals of uncertainty for the Mutagen dataset when the decision threshold is 0.7**

| interval of uncertainty | ridge logistic regression | | | random forest | | |
|---|---|---|---|---|---|---|
| | specificity | sensitivity | percent Uncertain | specificity | sensitivity | percent Uncertain |
| - | 0.691 | 0.875 | 0.00 | 0.557 | 0.933 | 0.00 |
| (0.69 – 0.71) | 0.696 | 0.872 | 1.39 | 0.561 | 0.930 | 2.11 |
| (0.68 – 0.72) | 0.701 | 0.869 | 2.76 | 0.565 | 0.927 | 4.43 |
| (0.67 – 0.73) | 0.706 | 0.866 | 4.16 | 0.569 | 0.923 | 6.75 |
| (0.66 – 0.74) | 0.711 | 0.863 | 5.49 | 0.573 | 0.919 | 9.00 |
| (0.65 – 0.75) | 0.716 | 0.860 | 6.89 | 0.578 | 0.914 | 11.50 |
| (0.64 – 0.76) | 0.721 | 0.856 | 8.31 | 0.582 | 0.910 | 13.74 |
| (0.63 – 0.77) | 0.727 | 0.852 | 9.74 | 0.586 | 0.905 | 15.67 |
| (0.62 – 0.78) | 0.733 | 0.848 | 11.13 | 0.591 | 0.900 | 17.78 |
| (0.61 – 0.79) | 0.739 | 0.844 | 12.57 | 0.596 | 0.894 | 19.84 |
| (0.60 – 0.80) | 0.745 | 0.839 | 14.01 | 0.602 | 0.888 | 21.91 |

Mean specificity and sensitivity without Uncertain predictions and mean percentage of Uncertain predictions are shown for the best ridge logistic regression model and random forest model on the Mutagen dataset when a decision threshold is 0.7.

**Table 6b - Analysis of intervals of uncertainty for the Mutagen dataset when the decision threshold is 0.3**

| interval of uncertainty | ridge logistic regression | | | random forest | | |
|---|---|---|---|---|---|---|
| | specificity | sensitivity | percent Uncertain | specificity | sensitivity | percent Uncertain |
| - | 0.919 | 0.587 | 0.00 | 0.964 | 0.408 | 0.00 |
| (0.29 – 0.31) | 0.917 | 0.593 | 1.53 | 0.963 | 0.412 | 2.02 |
| (0.28 – 0.32) | 0.916 | 0.599 | 2.99 | 0.962 | 0.416 | 4.24 |
| (0.27 – 0.33) | 0.914 | 0.606 | 4.51 | 0.961 | 0.418 | 5.84 |
| (0.26 – 0.34) | 0.912 | 0.612 | 6.06 | 0.960 | 0.422 | 7.87 |
| (0.25 – 0.35) | 0.910 | 0.619 | 7.62 | 0.958 | 0.426 | 9.90 |
| (0.24 – 0.36) | 0.908 | 0.626 | 9.19 | 0.957 | 0.430 | 11.96 |
| (0.23 – 0.37) | 0.906 | 0.633 | 10.76 | 0.955 | 0.434 | 14.01 |
| (0.22 – 0.38) | 0.904 | 0.641 | 12.44 | 0.953 | 0.437 | 16.07 |
| (0.21 – 0.39) | 0.902 | 0.648 | 14.00 | 0.951 | 0.442 | 18.10 |
| (0.20 – 0.40) | 0.899 | 0.656 | 15.68 | 0.949 | 0.446 | 20.16 |

Mean specificity and sensitivity without Uncertain predictions and mean percentage of Uncertain predictions are shown for the best ridge logistic regression model and random forest model on the Mutagen dataset when a decision threshold is 0.3.

**Table 7a - Analysis of intervals of uncertainty for the PLD dataset when the decision threshold is 0.7**

| interval of uncertainty | ridge logistic regression | | | random forest | | |
|---|---|---|---|---|---|---|
| | specificity | sensitivity | percent Uncertain | specificity | sensitivity | percent Uncertain |
| - | 0.262 | 0.993 | 0.00 | 0.374 | 0.983 | 0.00 |
| (0.69 – 0.71) | 0.266 | 0.993 | 3.05 | 0.379 | 0.982 | 2.27 |
| (0.68 – 0.72) | 0.269 | 0.993 | 6.10 | 0.384 | 0.981 | 5.03 |
| (0.67 – 0.73) | 0.273 | 0.992 | 9.12 | 0.390 | 0.981 | 8.17 |
| (0.66 – 0.74) | 0.278 | 0.992 | 12.01 | 0.398 | 0.980 | 11.22 |
| (0.65 – 0.75) | 0.283 | 0.991 | 15.21 | 0.405 | 0.979 | 14.38 |
| (0.64 – 0.76) | 0.288 | 0.991 | 18.40 | 0.411 | 0.978 | 17.40 |
| (0.63 – 0.77) | 0.294 | 0.990 | 21.69 | 0.417 | 0.976 | 20.07 |
| (0.62 – 0.78) | 0.300 | 0.990 | 24.77 | 0.424 | 0.975 | 23.15 |
| (0.61 – 0.79) | 0.308 | 0.989 | 28.31 | 0.431 | 0.974 | 25.99 |
| (0.60 – 0.80) | 0.317 | 0.988 | 32.00 | 0.440 | 0.972 | 28.99 |

Mean specificity and sensitivity without Uncertain predictions and mean percentage of Uncertain predictions are shown for the best ridge logistic regression and random forest model on the PLD dataset when a decision threshold is 0.7.

**Table 7b - Analysis of intervals of uncertainty for the PLD dataset when the decision threshold is 0.3**

| interval of uncertainty | ridge logistic regression | | | random forest | | |
|---|---|---|---|---|---|---|
| | specificity | sensitivity | percent Uncertain | specificity | sensitivity | percent Uncertain |
| - | 0.883 | 0.628 | 0.00 | 0.867 | 0.631 | 0.00 |
| (0.29 – 0.31) | 0.878 | 0.629 | 1.51 | 0.863 | 0.634 | 1.35 |
| (0.28 – 0.32) | 0.874 | 0.630 | 2.98 | 0.858 | 0.636 | 2.78 |
| (0.27 – 0.33) | 0.869 | 0.631 | 4.38 | 0.855 | 0.637 | 3.83 |
| (0.26 – 0.34) | 0.863 | 0.632 | 5.99 | 0.850 | 0.640 | 5.31 |
| (0.25 – 0.35) | 0.856 | 0.633 | 7.55 | 0.844 | 0.643 | 6.76 |
| (0.24 – 0.36) | 0.849 | 0.635 | 9.14 | 0.839 | 0.645 | 8.07 |
| (0.23 – 0.37) | 0.842 | 0.637 | 10.71 | 0.833 | 0.648 | 9.48 |
| (0.22 – 0.38) | 0.834 | 0.639 | 12.19 | 0.826 | 0.651 | 10.86 |
| (0.21 – 0.39) | 0.825 | 0.640 | 13.68 | 0.821 | 0.654 | 12.05 |
| (0.20 – 0.40) | 0.817 | 0.642 | 15.14 | 0.813 | 0.657 | 13.54 |

Mean specificity and sensitivity without Uncertain predictions and mean percentage of Uncertain predictions are shown for the best ridge logistic regression and random forest model on the PLD dataset when a decision threshold is 0.3.

**Table 8a - Analysis of intervals of uncertainty for the PLD dataset when the decision threshold is 0.7**

| interval of uncertainty | ridge logistic regression | | | random forest | | |
|---|---|---|---|---|---|---|
| | specificity | sensitivity | percent Uncertain | specificity | sensitivity | percent Uncertain |
| - | 0.596 | 0.900 | 0.00 | 0.635 | 0.891 | 0.00 |
| (0.69 – 0.71) | 0.598 | 0.897 | 1.24 | 0.637 | 0.888 | 1.34 |
| (0.68 – 0.72) | 0.598 | 0.894 | 2.43 | 0.639 | 0.883 | 3.22 |
| (0.67 – 0.73) | 0.599 | 0.891 | 3.80 | 0.641 | 0.878 | 4.96 |
| (0.66 – 0.74) | 0.599 | 0.888 | 4.94 | 0.646 | 0.873 | 6.68 |
| (0.65 – 0.75) | 0.600 | 0.884 | 6.25 | 0.649 | 0.868 | 8.18 |
| (0.64 – 0.76) | 0.602 | 0.879 | 7.90 | 0.653 | 0.864 | 9.70 |
| (0.63 – 0.77) | 0.603 | 0.875 | 9.06 | 0.656 | 0.859 | 11.19 |
| (0.62 – 0.78) | 0.605 | 0.872 | 10.13 | 0.658 | 0.852 | 12.71 |
| (0.61 – 0.79) | 0.607 | 0.867 | 11.57 | 0.660 | 0.845 | 14.33 |
| (0.60 – 0.80) | 0.608 | 0.862 | 12.86 | 0.665 | 0.838 | 16.00 |

Mean specificity and sensitivity without Uncertain predictions and mean percentage of Uncertain predictions are shown for the best ridge logistic regression and random forest model on the bbb2 dataset when a decision threshold is 0.7.

**Table 8b  - Analysis of intervals of uncertainty for the PLD dataset when the decision threshold is 0.3**

| interval of uncertainty | ridge logistic regression | | | random forest | | |
|---|---|---|---|---|---|---|
| | specificity | sensitivity | percent Uncertain | specificity | sensitivity | percent Uncertain |
| - | 0.954 | 0.331 | 0.00 | 0.960 | 0.495 | 0.00 |
| (0.29 – 0.31) | 0.951 | 0.336 | 3.59 | 0.958 | 0.499 | 2.58 |
| (0.28 – 0.32) | 0.949 | 0.341 | 6.61 | 0.956 | 0.507 | 4.91 |
| (0.27 – 0.33) | 0.946 | 0.345 | 9.54 | 0.955 | 0.509 | 6.56 |
| (0.26 – 0.34) | 0.943 | 0.348 | 12.56 | 0.953 | 0.515 | 9.09 |
| (0.25 – 0.35) | 0.940 | 0.351 | 15.37 | 0.952 | 0.521 | 11.44 |
| (0.24 – 0.36) | 0.936 | 0.355 | 18.48 | 0.949 | 0.527 | 14.08 |
| (0.23 – 0.37) | 0.931 | 0.359 | 21.92 | 0.947 | 0.531 | 15.95 |
| (0.22 – 0.38) | 0.926 | 0.364 | 24.58 | 0.945 | 0.539 | 18.46 |
| (0.21 – 0.39) | 0.920 | 0.369 | 27.82 | 0.942 | 0.547 | 20.96 |
| (0.20 – 0.40) | 0.913 | 0.373 | 30.91 | 0.939 | 0.553 | 23.52 |

Mean specificity and sensitivity without Uncertain predictions and mean percentage of Uncertain predictions are shown for the best ridge logistic regression and random forest model on the bbb2 dataset when a decision threshold is 0.3.